# Intuitive Source Code Visualization Tools for Improving Student Comprehension: BRICS

Christopher Pearson, Celina Gibbs, Yvonne Coady
University of Victoria
pearson@csc.uvic.ca, celinag@cs.uvic.ca, ycoady@cs.uvic.ca

## ABSTRACT
Even relatively simple code analysis can be a daunting task for many first year students. Perceived complexity, coupled with foreign and harsh syntax, often outstrips the ability for students to take in what they are seeing in terms of their *verbal memory*. That is, first year students often lack the experience to encode critical building blocks in source code, and their interrelationships, into their own words. We believe this argues for the need for IDEs to provide additional support for representations that would appeal directly to *visual memory*. In this paper, we examine this need for intuitive source code visualization tools that are easily accessible to novice programmers, discuss the requirements for such a tool, and suggest a novel idea that takes advantage of human peripheral vision to achieve stronger overall code structure awareness.

## 1. INTRODUCTION
Students who are new to software development face many hurdles, from thinking logically to understanding objects. During this period, many are overloaded with having to learn the functionality of development environments that are overly complicated for their needs. There are several good examples of trimmed down development environments that are designed with the CS1 student in mind, such as GILD [1], jGRASP [2] and BlueJ [3]. GILD works to bring pedagogical support to the Eclipse environment. Both jGRASP and BlueJ are stand alone IDEs that present simplified UML class diagrams. These allow the student to experiment with class relations without needing to understand the nuances of full UML class diagrams. Further work exists to present minimalistic views in Eclipse to reduce the initial learning curve [4].

Regardless of whether a CS1 introductory course is taught as objects-first, students must still go through the process of learning to understand source code. There have been many improvements since the days of the simple text editor, such as code indentation and syntax highlighting, and both of these developments have helped students with their understanding of source code. However, with one exception, the evolution of code display seems to have come to a halt.

That exception is the jGRASP environment, which introduces the Control Structure Diagram (CSD) - a method of drawing vertical bars to the left of the source code to highlight the structure of the control blocks (See Figure 1). The need for the CSD is described well by Hendrix et al. [5]:

> Visualizations such as flowcharts disrupt the layout of the source code by viewing a program as a flow graph with source code statements or statement fragments attached to nodes of the graph. Thus a reader is forced to comprehend the source code through a completely different notation and layout (i.e., the flowchart). The CSD appears as a companion to rather than a replacement for source code, thus leveraging the perceived advantages of a graphical representation together with the familiarity of pretty-printed source code.

In this paper, Hendrix et al. also present convincing evidence for the effectiveness of their method.

When CS1 students are initially taught source code control flow, diagrams are often used for the explanations. Blocks are drawn to represent the functional pieces of a program, giving the student an overview of how the code interacts. Increasingly larger blocks can be found from the smallest control structures through to methods and classes and packages and systems. At present, with the exception of jGRASP's CSD, we are not aware of any available tools to assist the student with the visualization of these simple structures.

## 2. VISUALIZATION
This research is inspired by works on visualizing information, as well as works on how the human brain handles text and images. Specifically, we are interested in research suggesting that multiple views of data can be more effective than single views with zoom functions [6], and that people naturally focus on visual objects when highlighting cues are present [7]. Also of interest is research that describes visual working memory as being separate from verbal working memory, and demonstrates that each has their own small capacity for holding information [8]. As the visual memory applies to shapes and the verbal memory applies to text, and as both types of working memory can only hold a very limited number of items, we would suggest that the combination of source code with diagrams would be a powerful one.

We have noted that while many tools for visualizing objects and data structures have appeared in recent years, the visualizing of source code itself seems to have been forgotten. To a large extent, people work visually, and we should be taking greater advantage of this when teaching source code flow.

## 3. TOOL REQUIREMENTS
While developing our vision of intuitive diagramming software, a list of requirements for such a tool was created. These are the features we feel are required for such a tool to be successful.

### 3.1 Multiple Views
Multiple views should be supported, with a graphical overview window available to supplement any diagrams. As discussed by Plumlee et al. [6], when trying to keep the overall picture in mind, the human eye will frequently look to a graphical overview if one is available. This follows cognitive research showing that the limits of short term memory are quite small [8], and that the brain will do its best to use any well designed supplemental information. It then follows that a visual display tool should be accompanied by an overview window to give reference and context.

Regardless of the alternate views, the students must learn to interpret the code line by line, and so a reasonably sized source code window should always be available to give the student an opportunity to directly compare the views. The visual tool's abstract model may assist the student in learning the structure at a faster pace, while the availability of the source code's concrete view helps to ensure that the relationship between the two models is understood with clarity.

### 3.2 View Flexibility
If an IDE allows for custom views, it is important to ensure that components of the tool can be moved and reshaped according to the user's wishes, all while remaining functional.

With the rapid growth in multiple monitor systems, it is important to allow for parts of the view to be "pulled off" and to become free floating windows. A current example of this functionality can be seen in Eclipse where such tabs as console or help can be removed from the main window and dragged to another monitor.

### 3.3 Intuitive and Nonintrusive
Visual additions to development environments typically add some form of interface clutter. A fine example of this can be seen in Eclipse when many plug-ins are active. Rather than continually adding features to interfaces, we believe an important goal should be to limit the number of visual objects that the student needs to comprehend.

Ideally, such a tool should be intuitive enough that it results in a reduction in the amount of work that is required of a student learning to interpret code structure. It would be a tool that is used by the student from the start of the project, and not one that is used after the fact to generate the diagrams that were required for an assignment.

The tool should also integrate well with the development environment and not interfere with other features. During our comparisons of existing Eclipse plug-ins with visualizations, we found that some required their own configurations of the views, and that choosing to use such tools would force a change in the developer's personal preferences.

### 3.4 Synchronization and Interactivity
An attribute that is lacking in many current visualization tools is interactivity. Some diagramming tools are purely for the purpose of creating the diagram and do not synchronize with the source code. We believe that students need tools that allow for manipulation of code through diagram interaction. The features in the diagrams should be active and respond to user input.

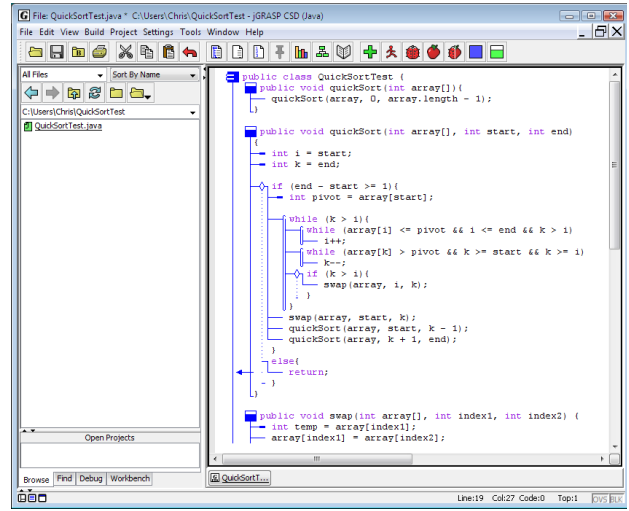

**Figure 1. The Control Structure Diagram (CSD) presented by jGRASP.**

Tools that are written for CS1 students should have a fast and automatic synchronization between the source code and any visual representations. Immediate updates to the visualization ensure that the student cannot become confused due to conflicting information in different views. As an example, the CSDs in the current version of jGRASP (1.8.6_03) require a button to be pushed for a visualization update to occur, and as such the CSD structure lines can quickly become out of sync during editing.

Synchronization must also be very reliable. If a visual model occasionally fails to update, it would cause great amounts of confusion. This is a problem we have seen while testing diagramming tools with our students.

### 3.5 Usability Beyond CS1/CS2
A truly useful tool is one that students would continue to use beyond their introduction to programming. Preferably it would be a tool that we ourselves would use regularly. Although cases can be made for tools that are constructed solely for the purpose of helping students understand code structure, it is reasonable to consider constructing a tool that the student can continue to use beyond CS1/CS2. In a profession that promotes code reuse, it seems to make sense to consider tool reuse.

If we are to consider a tool that will continue to be useful, we must naturally consider the environment it is designed for. A tool that is built as part of a non-extensible environment would not be of use further in the student's career. For this reason alone we believe the trials of such a tool should take place within an extensible professional development environment such as Eclipse. It should be noted that BlueJ, while designed as a learning environment, does have good support for extension.

## 4. A SOLUTION: BRICS
Our solution includes an extension to the CSD diagram that is designed to be nonintrusive and to take advantage of the human eye's peripheral vision. To do so we replace the vertical lines of the CSD with subtle boxes that wrap and nest around each control structure. We refer to this method as Blocks of Rationally Intuitive Control Structures (BRICS).

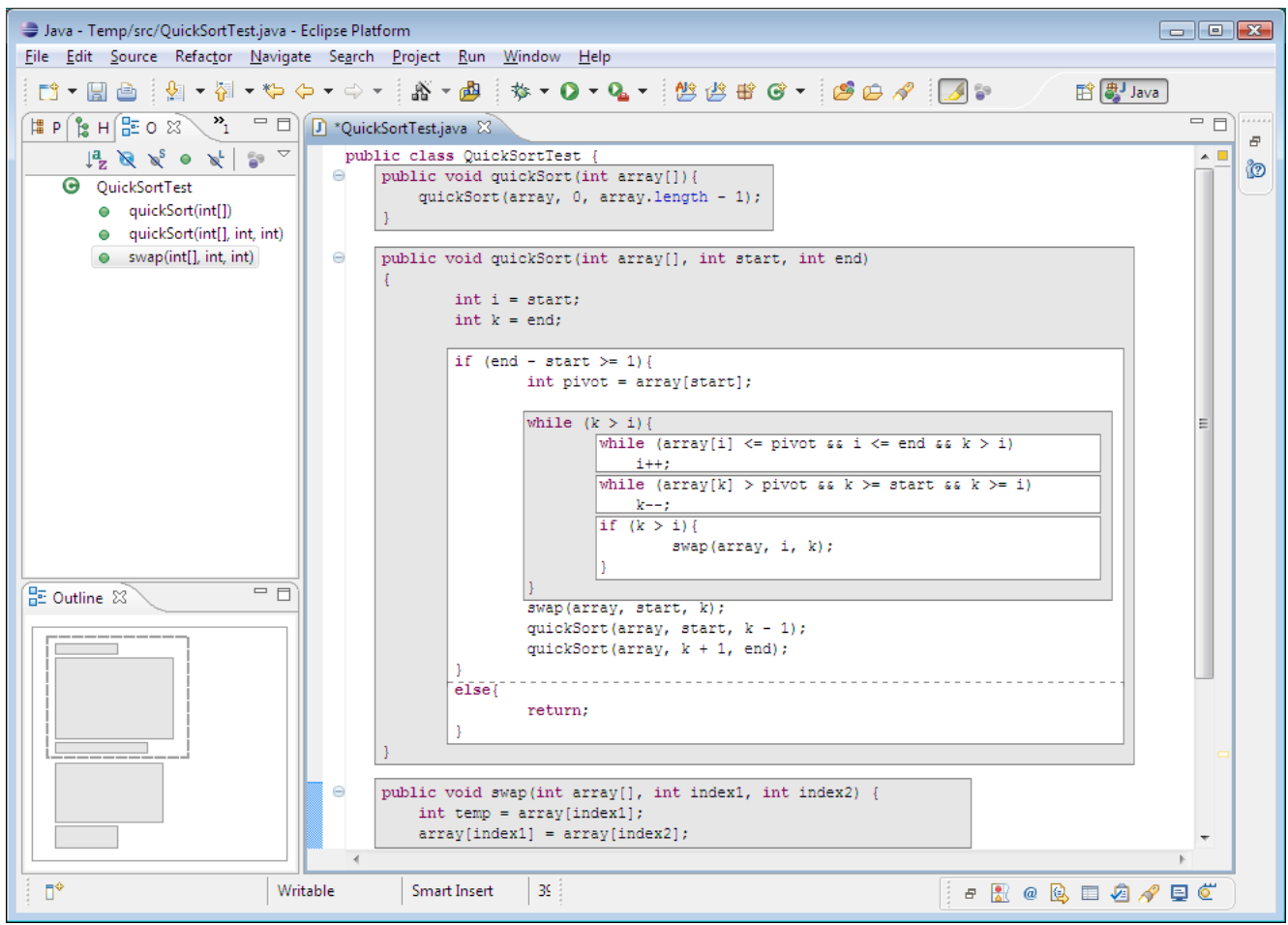

**Figure 2. The vision of BRICS as an Eclipse plug-in, showing the source code and overview windows.**

As being nonintrusive is one of the important requirements, BRICS is currently envisioned as undecorated boxes with alternating subtle neutral coloring. (See Figure 2) The intention is for them to remain in the background and to not interfere in any way with existing text highlighting. It appears that a slightly darker outline assists in seeing these rectangular shapes, and there is some research that seems to support this [9].

The important advantage of this technique, and the one we believe best shows the true power of BRICS, is that it takes advantage of the human eye's peripheral vision. When the human eye is focusing on a specific point, the brain will see any simple shapes that are nearby. As such, while the student is reading through the source code, they would be able to identify the overall structure of the surrounding code. Without these simple shapes neither code indentation nor jGRASP's CSD (see Figure 2) can give an equally strong intuitive effect. We believe this could be a powerful method for assisting the CS1 student grasp the concepts of structure and flow.

Having drawn the boxes around the control structures, we begin to have a truly visual understanding of the flow of the code in a similar manner to that of a flow chart. What is important, however, is the fact that we are giving that visual nature in combination with the actual source code and reinforcing the connection between the two. BRICS blends an abstract model view with the concrete code view to achieve a result that is, we believe, greater than the sum of its parts.

## 4.1 Overview Windows

Once a graphical representation of the source code exists, the creation of an overview can happen as in any other diagramming tool. The overview would retain the same ratio of the block shapes seen in the source code window to keep the student oriented with its context. Having an overview window that shows more of the BRICS structure can only help to assist in the understanding of the flow of the program. With these two windows the student would find a large amount of information regarding proximity, locality, and scope.

To allow for personal choice, the BRICS overview window requires granularity controls to choose the level of block nesting to be shown. Zoom controls would allow the user to choose how much of the current source file is being shown, and thereby explore the special relationships between various code structures.

Many other interesting possibilities exist, an example of which is integration with other editor features. Perhaps showing red lines in the overview to match compiler errors would be useful. Coloring the blocks depending on JUnit test results, or degree-of-interest information from Mylyn, may be worth testing.

## 4.2 Manipulating code with BRICS

With the concept of BRICS in place, we turned to the consideration of what this could be used for, beyond source code visualization, without adding complexity. Some ideas stood out, such as refactoring and code collapsing.

As an example, the Eclipse environment provides a tool which will take a selected block of code and refactor it into a separate method, all while maintaining dependencies. It should be a reasonable next step to allow the user to grab a block and drag it out of its method. We imagine this being used in an introductory CS1 assignment where an average student has written a very large method full of control structures. Using BRICS, this student would clearly see what sections of the code are grouped together and could easily drag this code out to become a new method. This would be a powerful tool given the difficulties we have observed CS1 students experiencing when attempting the decomposition of their code into more manageable methods.

## 4.3 Usability Beyond the Classroom

Beyond the introductory software development classes, BRICS has the potential to be valuable to professional software developers as well. Our initial goal was to design tools that we would use ourselves, and BRICS has met that criterion. Of note are the self documenting nature of BRICS and the enhanced code traceability that is inherent in the design.

Any tool that can help to facilitate an easier understanding of lengthy source code is something any professional developer could appreciate.

## 4.4 BRICS and the Requirements

Having defined our vision of BRICS, we compare it to our list of requirements for such a tool. Table 1 shows how the design of BRICS addresses each one of the requirements established in Section 3. We believe that by integrating the view with the source and introducing minimal tradeoffs in clutter and performance bodes well for a tool that could help beginner programmers understand code. Additionally, we believe these principles will scale to more sophisticated programming constructs beyond CSDs and aid code analysis even in professional settings.

**Table 1. BRICS and the Requirements**

| Requirement | BRICS as a solution |
|---|---|
| Multiple Views | Combines the diagram view with the source code itself, negating the need for multiple views. |
| View Flexibility | Dependant on implementation environment. |
| Intuitive and Nonintrusive | Designed to be subtle and intuitive, avoiding adding interface clutter. |
| Synchronization and Interactivity | Interactive by design, given that it all interactivity is with the source code editor. |
| Usability Beyond CS1/CS2 | Highlights the structure of source code, which is of use to students and professional developers alike. |

## 5. FUTURE RESEARCH

BRICS currently exists only as a concept. There are many challenges ahead, starting with integrating the blocking with the source editor. Our intentions are to create a prototype of BRICS for the Eclipse environment, and a great deal of research will need to be done to understand the amount of work required to implement it.

Before a prototype exists, we intend to present sample code mock-ups to students at various levels to gain an understanding of how well BRICS would be received by novice and expert programmers alike. Our initial impromptu displays of BRICS have seen a very warm reception and, although these results are anecdotal, they would indicate that BRICS indeed has the positive impact we are anticipating.

We intend to look beyond BRICS as currently envisioned to discover what other intuitive visualizations can be added. Some additions being considered include outlining the blocks with Mylyn degree-of-interest colorings, as well as adding red lines to the overview window to represent the location of errors. It has been suggested that BRICS would be very useful when editing C source code if the functionality was available to color `#ifdef` blocks differently depending on whether they are to be compiled in or not.

The application of BRICS appears to be widespread and varied, and as such it will be developed in a way similar to current source code highlighting systems in that configuration files will define what constitutes a structure to be blocked. We wish to ensure that BRICS can be used with as many programming languages as possible.

## 6. CONCLUSION

Currently there is a lack of intuitive introductory level source code visualization tools. While there is no simple solution to this problem, there exists the potential for great improvement. If tools are designed to avoid adding to interface clutter while being simple to learn and use they would have a dramatic impact on how students learn to interpret source code.

We hope this work will inspire other researchers to look for simple yet effective visualization tools for teaching CS students in their first year, and beyond.

## 7. ACKNOWLEDGMENTS

We would like to give our thanks to Cara Blanchard, Torben Eisler, Elizabeth Gasten, Katherine Gunion, and Onat Yazir for their support and suggestions.